\documentclass[a4paper,11pt]{article}
\pdfoutput=1

\usepackage{jheppub} 

\usepackage[T1]{fontenc} 
\usepackage{xcolor}

\usepackage{booktabs}
\usepackage{float}
\usepackage{titlesec}
\usepackage{capt-of}
\usepackage{slashed}

\usepackage{array}
\usepackage{arydshln}
\newcommand{\chitop}{\chi_{\rm top}}
\newcommand{\Qtop}{Q_{\rm top}}
\newcommand{\qtop}{q_{\rm top}}
\newcommand{\dd}{\textmd{d}}
\newcommand{\Z}{\mathcal{Z}}

\title{\boldmath Impact of extreme magnetic fields on the QCD topological susceptibility in the vicinity of the crossover region}

\author{B.B. Brandt,}
\author{G. Endr\H{o}di,}
\author{J.J. Hern\'andez Hern\'andez}
\author{and G. Mark\'o}
\affiliation{Universität Bielefeld,\\Universitätsstraße 25, 33615 Bielefeld, Germany}

\emailAdd{brandt@physik.uni-bielefeld.de}
\emailAdd{endrodi@physik.uni-bielefeld.de}
\emailAdd{hernandez@physik.uni-bielefeld.de}
\emailAdd{gmarko@physik.uni-bielefeld.de}

\abstract{We present the first determination of the topological susceptibility from lattice QCD in the presence of strong background magnetic fields. Our simulations employ 2+1 flavours of stout-improved staggered quarks with physical masses and cover a broad range of temperatures and magnetic field values. The results are extrapolated to the continuum limit  using four different lattice spacings and an eigenvalue reweighting technique to reduce discretisation errors. For low temperatures, our calculations show an enhancement of the topological susceptibility due to the magnetic field, compatible with predictions from chiral perturbation theory. At high temperatures, we observe the impact of inverse magnetic catalysis on the susceptibility.}

\begin{document} 
\maketitle
\flushbottom

\section{Introduction}

The vacuum of Quantum Chromodynamics (QCD) is a highly non-trivial state with inherently non-perturbative features and an intricate topological structure. Besides governing the prominent properties of the strongly interacting vacuum, topology also plays a key role in various features of QCD like chiral symmetry breaking, the anomalous breaking and effective restoration of the $\mathrm{U}(1)_{\rm A}$ symmetry, as well as the QCD phase diagram. 

Particular relevance is carried in these contexts by the amount of topological fluctuations, encoded by the topological susceptibility $\chitop$. Albeit being a purely QCD observable, the susceptibility also controls the behaviour of a hypothetical beyond-Standard-Model particle, the axion, which is a potential dark matter candidate and, simultaneously, provides a solution to the strong CP problem. 

Specifically, at zero temperature, the topological susceptibility explains the large mass of the $\eta'$ meson through the Witten-Veneziano formula~\cite{Witten:1979vv,Veneziano:1979ec}, and also fixes the normalisation of the quark condensate $\langle\bar\psi\psi\rangle$ in the chiral limit~\cite{DiVecchia:1980yfw,Colangelo:2010et}. In addition, the topological susceptibility is directly related to the mass of the axion -- thus at high temperatures, $\chitop$ carries information about the history of the axion field during its cosmological evolution~\cite{Preskill:1982cy,Abbott:1982af,Dine:1982ah}. In turn, axions might also play a key role in astrophysics, in particular for the anomalous cooling of neutron stars, see e.g.~\cite{Caputo:2024oqc,Sedrakian:2015krq}. In this last setting, the impact of nonzero density and background magnetic fields, pertinent to the discussion of neutron stars and magnetars, is of particular interest.
The coupling of axions to the electromagnetic field, is also crucial for the experimental detection of this hypothetical particle, see the recent review~\cite{Irastorza:2018dyq}.

Another setting where the topological features of the QCD medium and the effects of background magnetic fields intertwine in an intricate way is in off-central heavy-ion collisions. In this case, the chirality imbalance induced by topology and the background magnetic field produced by spectator particles lead to the chiral magnetic effect (CME)~\cite{Fukushima:2008xe}. The CME has been in the focus of experimental efforts for a long time, see the recent review~\cite{Kharzeev:2024zzm}. In this context, the impact of nonzero temperatures and background magnetic fields on the topological QCD fluctuations is also relevant, see e.g.~\cite{Asakawa:2010bu}.

The temperature-dependent topological susceptibility can be determined via different theoretical approaches, see e.g.\ the review~\cite{Vicari:2008jw}. Most importantly, lattice QCD simulations have been carried out both in the quenched approximation~\cite{Berkowitz:2015aua,Borsanyi:2015cka,Jahn:2018dke} as well as in the presence of physical dynamical quarks~\cite{Bonati:2015vqz,Petreczky:2016vrs,Borsanyi:2016ksw,Taniguchi:2016tjc,Athenodorou:2022aay}. Analytical methods to calculate $\chitop$ include chiral perturbation theory (ChPT), a systematically controlled expansion of QCD valid at low temperatures~\cite{GrillidiCortona:2015jxo}, as well as QCD models like the Nambu-Jona-Lasinio (NJL) model, see e.g.~\cite{Costa:2008dp,Ali:2020jsy}.

In this paper we focus on the impact of strong background magnetic fields on the topological susceptibility. This setup has been discussed in the literature so far using chiral perturbation theory~\cite{Adhikari:2022vqs,Adhikari:2021jff,Adhikari:2021lbl} and in the NJL model~\cite{Bandyopadhyay:2019pml}. On the lattice, the topological charge density correlator has been determined at low temperatures in the presence of magnetic fields~\cite{Bali:2013esa}. Notable lattice results about topology and magnetic fields include different indirect investigations of the chiral magnetic effect~\cite{Buividovich:2009wi,Abramczyk:2009gb,Buividovich:2009my,Braguta:2010ej,Bali:2014vja}. For a recent review on lattice QCD studies of the impact of magnetic fields, see~\cite{Endrodi:2024cqn}.
In summary, there are no existing lattice results for the combined dependence of $\chitop$ on magnetic fields and temperature, and this is our main goal in this work.

This paper is structured as follows. In Section~\ref{sec2} we briefly introduce the topological susceptibility in Euclidean space, its relation to the axion mass and our methodology. We describe the lattice setup of the simulations, the discretisations of the topological charge, and showcase that we can extract reliably the topology by using a smearing technique (the gradient flow) and explain how the reweighting of the fermion determinant affects the distribution of the topological charge. In Section~\ref{sec3} we present our results. Here we show how the eigenvalue spectrum on the lattice is different at low and high temperatures due to discretisation effects. We introduce a novel technique to solve the issue of misidentifying topological modes, which proves effective for the computation of ratios of topological susceptibilities. We also present the main result of the paper, namely the continuum extrapolation of the topological susceptibility as a function of temperature and the magnetic field. Finally in Section~\ref{sec4} we summarise our findings and give an outlook for future investigations. Our preliminary results have been presented previously in~\cite{Brandt:2022jfk,Brandt:2023awt}.

\section{The topological susceptibility and lattice methods}
\label{sec2}

\subsection{The topological susceptibility}

The topological susceptibility can be formally obtained from the QCD partition function $\Z$ defined through the Euclidean path integral, in the presence of a $\theta$ parameter. The latter is coupled to the topological charge $\Qtop$ in the Euclidean QCD action $S$ (see e.g.\ the review~\cite{Vicari:2008jw}),
\begin{equation}
S(\theta)=S - i\theta \,\Qtop, \qquad
\Qtop=\int \dd^4x \,\qtop(x), \qquad
\qtop(x)=\frac{g^2}{64\pi^2}\epsilon_{\mu\nu\rho\sigma}F^a_{\mu\nu}(x)F^a_{\rho\sigma}(x)\,,
\label{eq:Qtopqtop}
\end{equation}
where $F_{\mu\nu}^a$ is the gluon field strength, $g$ the strong coupling and we also defined the topological charge density $\qtop(x)$.

Using the functional dependence $\Z(\theta)$, the topological susceptibility is defined via
\begin{equation} \label{chitop_def}
    \chitop \equiv -\frac{1}{\Omega} \left.\frac{\partial^2 \log\Z(\theta)}{\partial \theta^2}\right\vert_{\theta = 0} = \int \dd^4x\, \langle \qtop(x)\qtop(0)\rangle = \frac{\langle \Qtop^2 \rangle}{\Omega}\,,
\end{equation}
where $\Omega$ is the space-time volume and $\langle.\rangle$ denotes the expectation value with respect to the partition function $\Z(\theta=0)$.
In Eq.~\eqref{chitop_def}, we used the translational invariance of the topological charge density correlator as well as the CP-symmetry of the system, ensuring $\langle \Qtop \rangle=0$. We note moreover that the above definition of $\chitop$ is valid both at zero and nonzero temperature.\footnote{For a discussion on the relationship between the topological susceptibility in Euclidean and Minkowskian space-times, see Ref.~\cite{Meggiolaro:1998bh}.} 

A homogeneous axion field $a$\footnote{Only in this section we use $a$ to refer to the axion field. In the rest of the text, it denotes the lattice spacing.} couples to gluons in the same manner as the $\theta$ parameter, allowing for the identification $\theta\equiv a/f_a$, where $f_a$ is the characteristic axion scale. The latter is, in principle, a free parameter of the theory, which can be constrained experimentally~\cite{Irastorza:2018dyq}. Treating $\Z(a)$ as the effective partition function for the axion field as a background field and using Eq.~\eqref{chitop_def}, we can see that the susceptibility is proportional to the square of the mass $m_a$ of the axion,
\begin{equation}
    \chi_{\rm top} = -\frac{f_a^2}{\Omega} \left.\frac{\partial^2 \log\Z(a)}{\partial a^2}\right\vert_{a = 0} = m_a^2f_a^2\,.
\end{equation}

This relation is exact and tells us that we can probe the axion mass through considering topological observables in pure QCD.
In our study, we will consider QCD with a homogeneous background magnetic field coupled to the fermions. The above definition for the topological susceptibility as well as its relation to the axion mass remains valid while taking into account the relevant, magnetic field-dependent partition function.

\enlargethispage{\baselineskip}

\subsection{Lattice setup}

In this study we have simulated QCD with $2+1$ flavours of rooted staggered quarks at the physical point with background magnetic fields at finite temperature. The partition function $\Z$ can be written using the Euclidean path integral over the gluon links $U$ as,
\begin{equation}
    \Z = \int \mathcal{D}U e^{-\beta S_g[U]}\prod_f\left(\det \left[\slashed{D}_f(U,q_fB) + m_f\right]^{1/4}\right),
\end{equation}
where the product in $f$ runs over the up, down and strange quarks and $\beta = 6/g^2$ denotes the inverse gauge coupling. 

The simulations were performed using lattices of geometry $N_s^3 \times N_t$, where $N_s \, (N_t)$ denotes the number of points in the spatial (temporal) direction. The physical volume of the lattice is then given by $V = (aN_s)^3$ and its temperature by $T = (aN_t)^{-1}$, with $a$ being the lattice spacing. We have generated ensembles in lattices of size $24^3\times6,\, 24^3\times8,\, 28^3\times10$ and $36^3\times12$, varying the inverse coupling in order to scan a range of temperatures around the crossover region, $T=112-212$ MeV, while performing a simultaneous scan for three different magnetic fields, $eB = 0,\, 0.5,\, 0.8$ GeV$^2$.

The gluonic action $S_g$ is the tree-level Symanzik discretisation, whereas the Dirac operator is discretised using staggered quarks with three times stout-smeared links. The Dirac operator depends both on the gluon links and on the magnetic field, which appears in the product with the charges of the quarks. The latter have been fixed to their physical values $q_u/2=-q_d=-q_s=e/3$, with $e>0$ the elementary electric charge. The masses of the quarks $m_f$ have been set to their physical values (assuming degenerate up and down quark masses) using the line of constant physics determined in Ref.~\cite{Borsanyi:2010cj}. The $1/4$ power of the fermion determinant arises due the rooting procedure~\cite{Sharpe:2006re}.

The magnetic field has been considered as a classical background field inside the Dirac operator, hence no dynamical photons were simulated. The electromagnetic potential $A_{\mu}$ was included in a similar fashion as the gluonic one, entering the Dirac operator as $\mathrm{U}(1)$ phases $u_{\mu,f}=\exp{(iaq_fA_{\mu})}$ that multiply the $\mathrm{SU}(3)$ gluonic links. The potential was chosen in such a way that it creates a homogeneous magnetic field pointing in the positive $z$ direction. Due to the geometry of our lattices and the periodic boundary conditions for the links, the magnetic field is quantised as $eB = 6\pi N_b/(aN_s)^2$, with $N_b \in \mathbb{Z}$ being the quantum flux number~\cite{Bali:2011qj}. We have to choose $N_b$ appropriately for each ensemble in order to maintain the physical field fixed. This implies that there is an uncertainty in the value of the magnetic field in physical units of around 3\%.\footnote{We note that the magnetic field dependence of $\chitop$ could also be determined via a Taylor-expansion around $B=0$. Such Taylor-expansions are possible for magnetic field profiles devoid of the flux quantization condition, see the review~\cite{Endrodi:2024cqn}. The leading-order dependence takes the form of a correlator of $\Qtop^2$ and the magnetic susceptibility~\cite{Bali:2020bcn}, a noisy observable. Since a considerable part of the required $B>0$ ensembles were already generated in Refs.~\cite{Bali:2011qj,Bali:2012zg,Bali:2014kia}, we decided to use the direct simulations at $B>0$.}

The exact parameters used in the simulations can be found in Appendix~\ref{tab_parameters}.

\subsection{Lattice definition of the topological charge and the gradient flow}
\label{Qtop_def}

The fundamental operator of our study is the topological charge $\Qtop$. On the lattice, this operator admits several possible discretised forms. We have employed two different lattice definitions for the local operator $\qtop(x)$ with different scalings to the continuum limit in order to study the systematic errors associated to the choice of discretisation. The simplest discretisation is the one consisting of the smallest closed gluonic loops (plaquettes) and has lattice artifacts of $\mathcal{O}(a^2)$. Moreover, we also consider an improved version containing larger Wilson loops chosen in such a way that the lattice artifacts of the lowest orders cancel perturbatively. In particular, we use the definition proposed in~\cite{Bilson-Thompson:2002xlt}, which improves the scaling to lattice artifacts of $\mathcal{O}(a^4)$. However, both discretisations still have lattice artifacts of order $\mathcal{O}(g^2a^2)$.  In the following we will refer to the former definition as regular, and the latter as improved.

For a given discretisation of the topological charge density $\qtop(x)$, the continuum definitions~\eqref{eq:Qtopqtop} can be simply translated to the lattice. Accordingly, we compute the topological charge and the susceptibility by 
\begin{equation}
\Qtop = a^4\sum_{x}\qtop(x), \qquad
    \chi_{\rm top} = \frac{\langle \Qtop^2\rangle}{N_s^3N_ta^4}. 
\end{equation}
where the sum runs over all lattice sites $x$.

Furthermore, in order to calculate the topological charge on a given configuration, we need to remove the ultraviolet fluctuations of the gauge fields. The method chosen for this task is the gradient flow~\cite{Luscher:2010iy}. In particular, we use the Wilson action~\cite{Wilson:1974sk} for the differential equations that define the gauge links at non-zero flow time $\tau_f$. For this choice the gradient flow can be also called Wilson flow. Observables constructed from the gauge links obtained for positive $\tau_f$ are expected to be renormalised and, thus, finite in the continuum limit~\cite{Luscher:2013cpa}. The gradient flow procedure may also be viewed as an averaging of the fields at each point over a domain with mean-squared radius $\sqrt{8\tau_f}$. Hence, only fluctuations with characteristic lengths larger than $\sqrt{8\tau_f}$ survive. While this greatly facilitates the determination of the (infrared) topological structure of a configuration, one also has to keep in mind that an excessive amount of flow will destroy the information stored in the gauge fields. At finite temperature, the amount of flow time should not exceed $\tau_f^{\rm max} = 1/\left(8T^2\right)$~\cite{Petreczky:2016vrs}.  

Far from these extremes, the topological charge is expected to be independent of $\tau_f$. Hence, an application of the gradient flow to our lattice fields should lead to plateaus as a function of the flow time for the topological charge and susceptibility. We have always made sure that a sufficient amount of flow time has been employed such that those plateaus are reached. In particular, following Ref.~\cite{Borsanyi:2016ksw}, all our observables have been defined at $\tau_f = \tau_f^{\rm max}$.

\begin{figure}[ht]
  \centering
  \begin{minipage}[ht]{0.49\textwidth}
    \includegraphics[width=\textwidth]{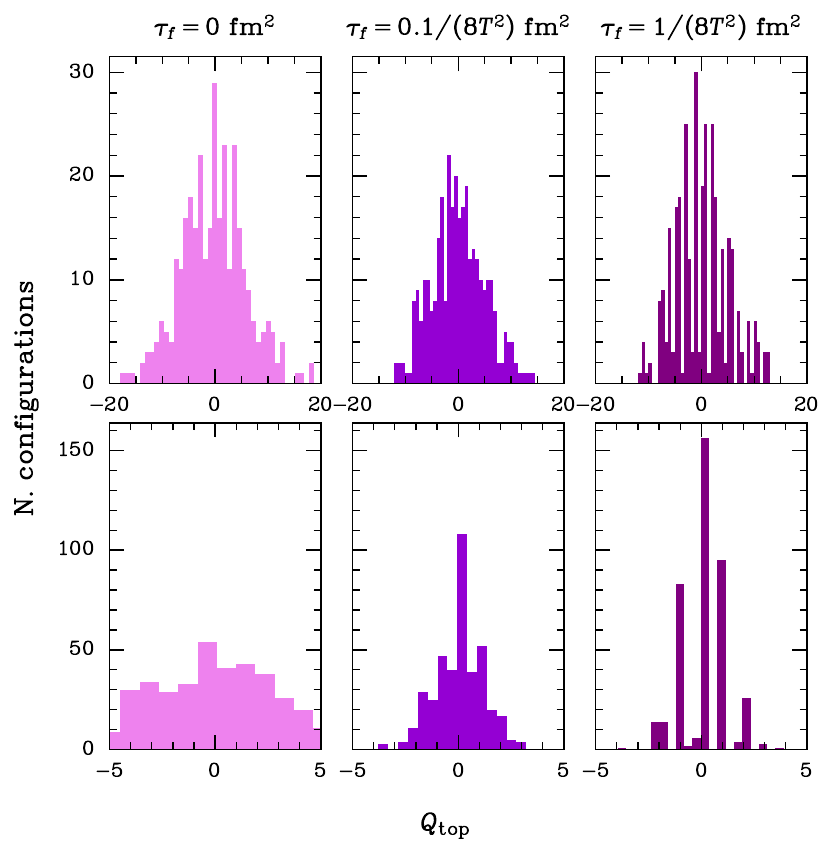}
  \end{minipage}
  \begin{minipage}[ht]{0.49\textwidth}
    \includegraphics[width=\textwidth]{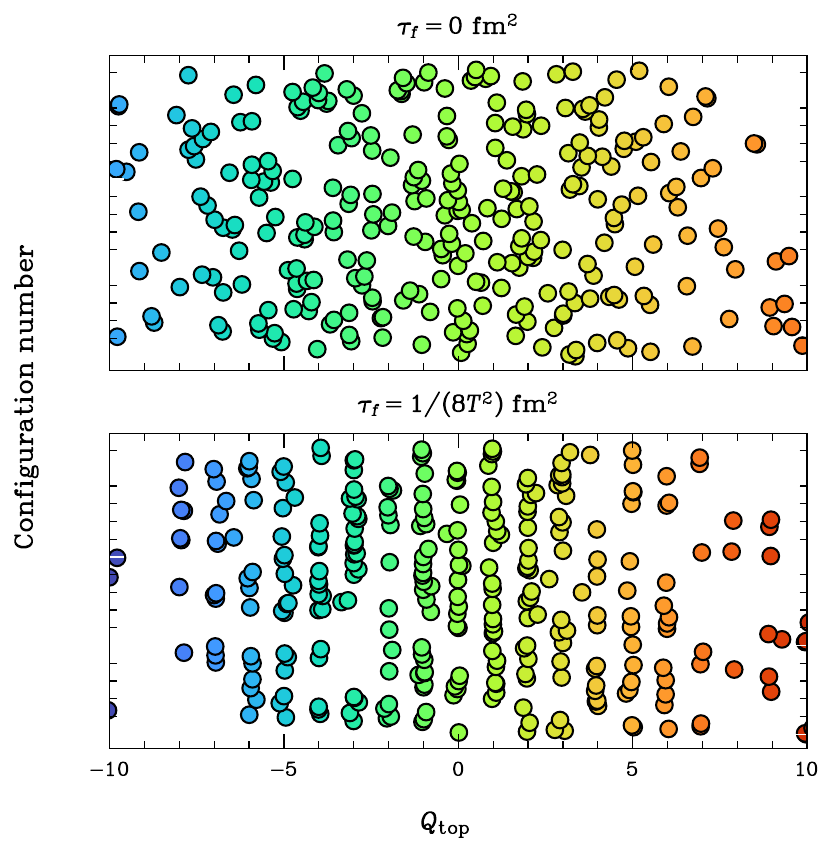}
  \end{minipage}
  \caption{\textsc{Left}: histograms of the topological charge on our $36^3\times 12$ lattice for three different values of the flow time (increasing from left to right), for two different temperatures (top, $T=150$ MeV; bottom, $T = 212$ MeV) at a nonzero magnetic field, $eB = 0.5$ GeV$^2$. \textsc{Right}: scatter plots of the topological charge for individual configurations in our $36^3\times 12$ ensemble at $T = 150$ MeV, at two different values of the flow time. We zoom in to the $[-10,10]$ window for visualisation purposes. Notice that $\Qtop$ approaches integers for almost all configurations. The improved definition of $\Qtop$ was used in all figures.}
  \label{histograms_flow}
\end{figure}

\begin{figure}[ht]
  \centering
    \includegraphics[width=0.5\textwidth]{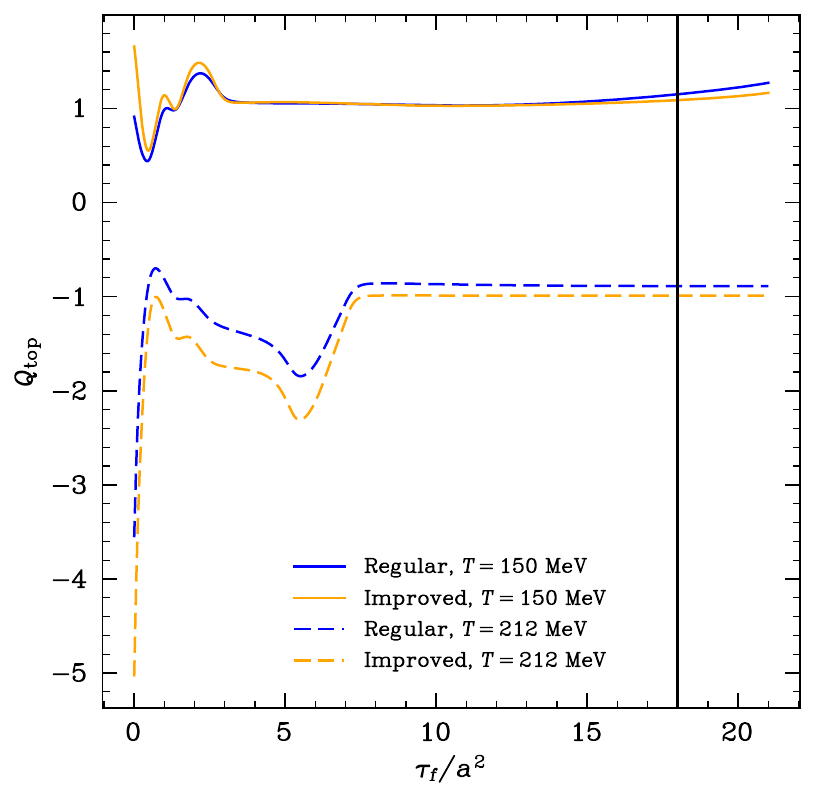}
  \caption{Evolution of the topological charge for two individual configurations under the Wilson flow, at two different temperatures, for the two discretisations used, with $eB = 0.5$ GeV$^2$, in our $36^3\times 12$ lattice. The dark vertical line denotes the instant where we extract the value of the operator ($\tau_f/a^2 = N_t^2/8$). We see how after a sufficient amount of flow time both discretisations are very close to integers.}
  \label{conf_flow}
\end{figure}

To demonstrate that this gradient flow technique is indeed under control, we show the effect of the Wilson flow on our $36^3\times 12$ lattices for two different temperatures, above and below the crossover temperature, namely for $T_1 = 150$ MeV and $T_2 = 212$ MeV, with a non-zero background magnetic field $eB = 0.5$ GeV$^2$, see Figs.~\ref{histograms_flow} and~\ref{conf_flow}. For the integration of the differential equation we have employed a fixed step size of $\Delta\tau_f/a^2 = 0.02$. 


Once we are close enough to the continuum limit, we are able to obtain almost integer values for the topological charge at all temperatures. We demonstrate this in the left panel of Fig.~\ref{histograms_flow}, where we plot the histograms of the topological charge at three different instants of the Wilson flow integration, obtaining almost perfect integers after some finite flow. This can be seen more clearly in the right panel of Fig.~\ref{histograms_flow}, where we compare the values of the topological charge for individual configurations before and after the integration. For completeness, we also show how the Wilson flow affects individual configurations, see Fig.~\ref{conf_flow}. 


Besides the two different discretisations of the topological charge described above, one can also use alternative definitions that stem from them, namely rounding the value of $\Qtop$ to the nearest integer or using a different value of $\tau_f$, i.e.\ flowing until 90\% of $\tau_f^{\rm max}$. This gives eight different definitions, which we studied in our analysis and considered the differences to the improved discretisation using rounding defined at $\tau_f = \tau_f^{\rm max}$.

\subsection{Reweighting of the fermion determinant}
\label{det_rw}

Even after employing the Wilson flow for $\Qtop$,
the topological susceptibility suffers from sizeable lattice artifacts that hinder a controlled continuum extrapolation, as shown by previous lattice calculations~\cite{Borsanyi:2016ksw}. Specifically, in our lattice setup $\chitop$ is about an order of magnitude larger than the expected continuum limit at our lowest temperature $T = 112\textmd{ MeV}$ and $B=0$~\cite{Brandt:2023awt}. Most of these lattice artifacts stem from the staggered discretisation of the Dirac operator. Due to the index theorem, on a gauge configuration with topological charge $\Qtop$, the Dirac operator possesses exactly $|\Qtop|$ many zero modes (in the continuum)\footnote{More precisely, this follows from the index theorem, ensuring that the difference of the numbers of zero modes with positive and negative chirality equals $\Qtop$ and the fact that in the absence of fine tuning, the Dirac operator only has zero modes with either positive or negative chirality (the so-called vanishing theorem in two dimensions~\cite{Nielsen:1977aw,Kiskis:1977vh}). For the lattice equivalent of the index theorem, see also Ref.~\cite{Hasenfratz:1998ri}.}. However, the staggered Dirac operator typically does not have exact zero modes. Hence, under the path integral, the weight of each configuration -- the determinant of the Dirac operator -- is overestimated for topological gauge fields. 

To correct for this effect, we introduce the reweighting factor~\cite{Borsanyi:2016ksw},
\begin{equation}
    W = \prod_{f}\prod_{j=1}^{4|Q_{\mathrm{top}}|}\prod_{\sigma=\pm1}\left(\frac{m_f}{\sigma i\lambda_{fj}+m_f}\right)^{1/4} = \prod_{f}\prod_{j=1}^{2|Q_{\mathrm{top}}|}\left(\frac{m_f^2}{\lambda_{fj}^2+m_f^2}\right)^{1/4}.
    \label{eq:Wdef}
\end{equation}

This is nothing but the ratio of the continuum fermion determinant and the staggered determinant in the topological sector, involving the exact zero modes and the corresponding staggered would-be zero modes.
In Eq.~\eqref{eq:Wdef}, $m_f$ is the mass, $\lambda_{fj}$ is the $j$-th lowest eigenvalue of the staggered Dirac operator for the quark flavour $f$ and $\Qtop$ is the topological charge, defined via the gluon fields and the Wilson flow, as explained above. The exponent $1/4$ arises due to the rooting procedure and the product in $j$ extends up to the correct number of zero-modes, which for the staggered operator is $4|\Qtop|$ owing to fermion doubling. Here we also took into account that the staggered eigenvalues come in complex conjugate pairs, and combined the pairs in the second step in Eq.~\eqref{eq:Wdef}.

Therefore, a better estimation of our observable can be obtained via reweighting, i.e.\ by multiplying each configuration by $W/\langle W\rangle$. This is equivalent to sampling the observable with an improved probability distribution, which is expected to facilitate the continuum extrapolation~\cite{Borsanyi:2016ksw}. 
Using this method, we can compute the expectation value of a reweighted observable as $\langle \mathcal{O}\rangle_{\mathrm{rw}}\equiv\langle W \mathcal{O}\rangle/\langle W \rangle$.
Notice that towards the continuum limit, the staggered would-be zero modes are expected to become exact zero modes~\cite{Durr:2005ax} so that $W$ approaches unity for all configurations. Thus, this kind of reweighting does not change the continuum limit, but it can reduce lattice discretisation errors substantially. The direct effects of the reweighting on the topological susceptibility at finite lattice spacings was reported in \cite{Brandt:2023awt}.

It is also worth noticing that in Eq.~\eqref{eq:Wdef} the would-be zero modes are identified with the lowest eigenvalues of the staggered Dirac operator. This identification is clear for configurations at high temperatures, where typical low-lying eigenvalues and would-be zero modes of topological origin are separated in the spectrum.
A physical interpretation is more complicated at low temperatures, where topological modes and near-zero modes building up the chiral condensate  mix. We will get back to this point below in Sec.~\ref{window}.

We also remark that the larger $|\Qtop|$, the smaller the reweighting factor, which suppresses the corresponding configuration. Hence, the reweighting in general reduces the topological susceptibility. In turn, for observables that are not directly sensitive to the topological modes, the above reweighting is not expected to have a significant effect. We demonstrate this for the quark condensate $\bar\psi\psi_f$ in Appendix.~\ref{rw_psibarpsi}. 

\begin{figure}[hb]
  \centering
    \includegraphics[width=\textwidth]{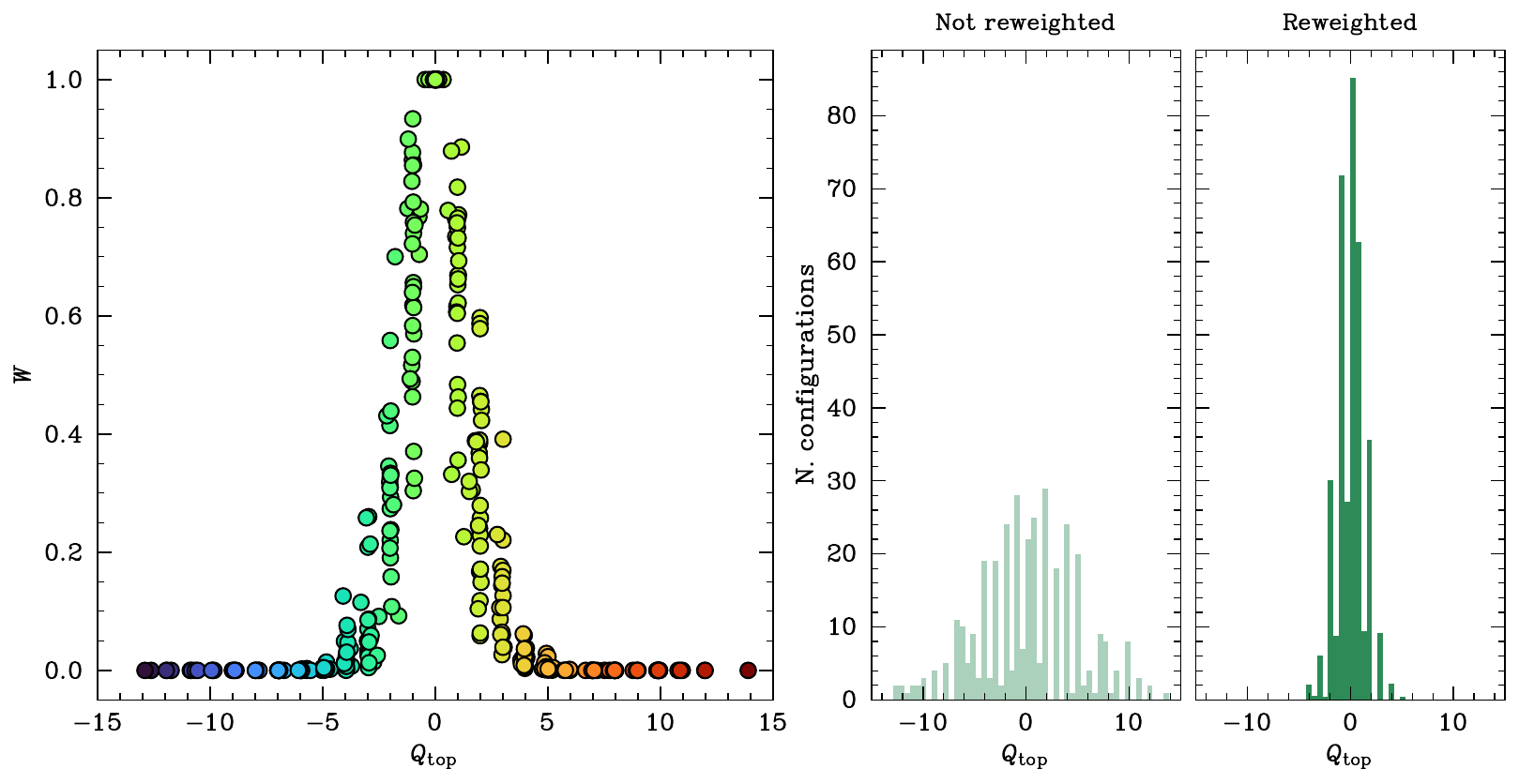}
  \caption{The ensemble under consideration is our $36^3\times 12$ lattice, at $T = 150 $ MeV and with a magnetic field of $eB = 0.8$ GeV$^2$. \textsc{Left}: scatter plot of the reweighting factor versus the topological charge. \textsc{Right}: histograms of the topological charge before and after reweighting. The improved definition of $\Qtop$ at a flow time instant of $\tau_f/a^2 = N_t^2/8$ was employed for both panels.}
  \label{effect_rw}
\end{figure}

The effect of the reweighting is shown in Fig.~\ref{effect_rw} for our $36^3\times12$ lattice, at a temperature of $T=150$ MeV and with a background magnetic field of $eB = 0.8$ GeV$^2$. On the scatter plot in the left panel we can observe how for large values of $|\Qtop|$, the reweighting factor highly suppresses the corresponding configurations. In turn, the contribution of configurations with low topology is effectively enhanced by the reweighting. In the right panel of Fig.~\ref{effect_rw}, we plot the histogram of the topological charge with and without reweighting, where it can be clearly seen how the width of the distribution (the topological susceptibility) is greatly reduced. We note  that alternative approaches to reduce staggered lattice artefacts include the use of fermionic definitions for $\Qtop$ like spectral projectors~\cite{Athenodorou:2022aay} as well as different types of smearing of the link variables, see e.g.~\cite{Bonati:2014tqa,Alexandrou:2017hqw}.

\section{Results}
\label{sec3}
\subsection{Eigenvalue spectrum}
\label{spectrum}

As commented in the previous section, the low end of the spectrum of the staggered Dirac operator is affected by lattice artifacts. We can distinguish two different regimes for these artifacts, based on the degree of separation between topological and non-topological modes as well as their chirality. The latter is defined\footnote{In fact, for our staggered formulation, the correct discretisation of $\gamma_5$ is the taste singlet matrix $\Gamma_5$, see the definition in e.g.\ Refs.~\cite{Durr:2013gp,Brandt:2024wlw}.} via the matrix element $\chi_{fj}^\dagger\gamma_5\chi_{fj}$ for the eigenmode $\chi_{fj}$ corresponding to the eigenvalue $\lambda_{fj}$. 

\begin{figure}[ht]
  \centering
    \includegraphics[width=\textwidth]{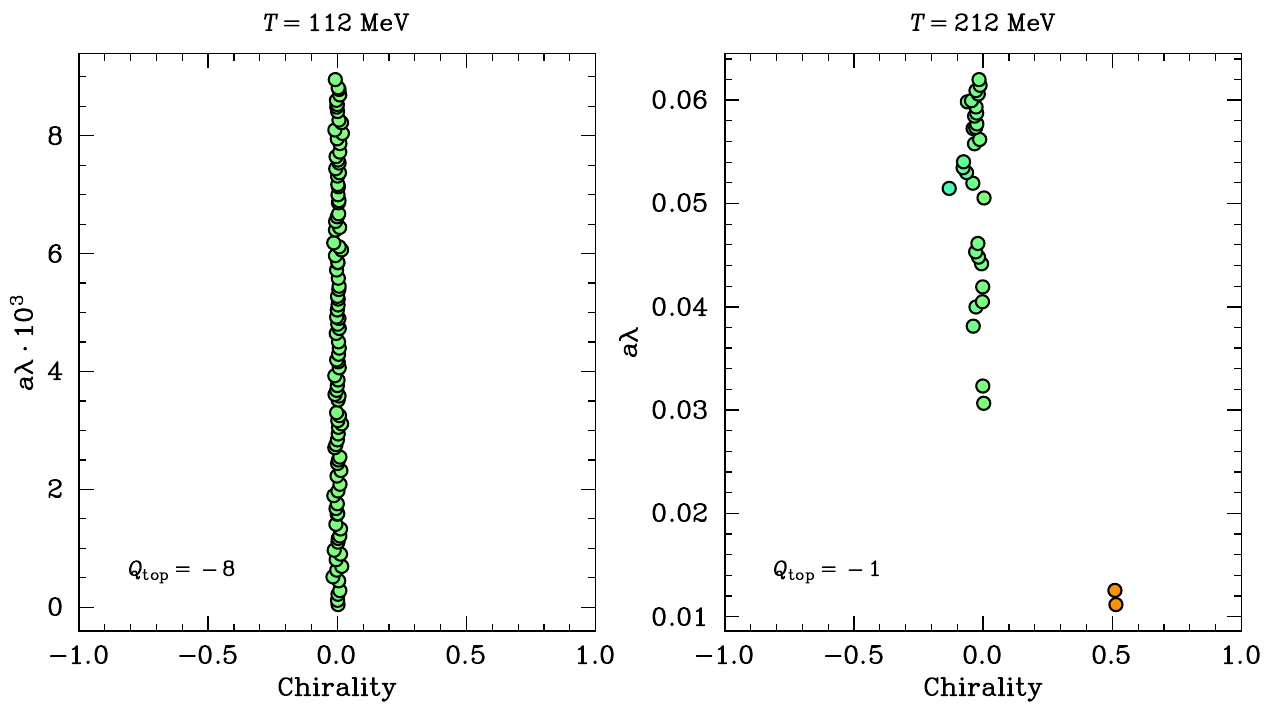}
  \caption{Scatter plots of the magnitude of the eigenvalue versus the chirality of the corresponding mode of the Dirac operator for the down (or strange) quark, in our $36^3\times12$ lattice, for two different temperatures, with a background magnetic field of $eB = 0.8$ GeV$^2$. Since the eigenvalues come in purely imaginary complex conjugate pairs, we only show the positive low-end of the spectrum. The value of $\Qtop$ was determined using the improved definition at $\tau_f/a^2 = N_t^2/8$.}
  \label{topo_separation}
\end{figure}

At low temperatures, the would-be zero modes of topological origin get mixed with the low-lying eigenvalues that build up the chiral condensate according to the Banks-Casher relation~\cite{Banks:1979yr}. As the temperature is increased, these low-lying modes disappear and only the topological would-be zero modes remain. We illustrate this picture in Fig.~\ref{topo_separation}. In the left panel of the figure, we see that at low temperatures there is no separation between the low-lying eigenvalues and the topological would-be zero modes, neither in magnitude nor in chirality. However, we see how this situation completely reverses at high temperatures (right panel of Fig.~\ref{topo_separation}) where now the eigenvalues of topological origin are clearly separated from the rest, both in magnitude and in chirality. Accordingly, in this setup one can clearly identify the value of $\Qtop$ and compare it with the gluonic definition, yielding the same result (as the index theorem dictates). The chirality of the would-be zero modes is not exactly $\pm 1$ due to discretisation effects.

\subsection{Multi-range random reweighting}
\label{window}

A clear identification of the eigenvalues of topological origin is necessary in order to reweight the fermion determinant, since those modes are the ones that we know become zero in the continuum. As shown in Fig.~\ref{topo_separation}, at high temperatures the identification is clear. However, this becomes problematic at low temperatures, since the would-be zero modes of topological origin may not coincide with the $4|\Qtop|$ smallest eigenvalues. In order to understand the error introduced by this ambiguity, we have studied the effect of reweighting not the $4|\Qtop|$ smallest eigenvalues but $4|\Qtop|$ randomly selected eigenvalues from a window in the spectrum that contains the ${\rm round}(W_F \cdot 4|\Qtop|)$ lowest-lying modes with $W_F>1$. We refer to this procedure as multi-range random reweighting and will call $W_F$ the window factor.

\begin{figure}[ht]
  \centering
    \includegraphics[width=\textwidth]{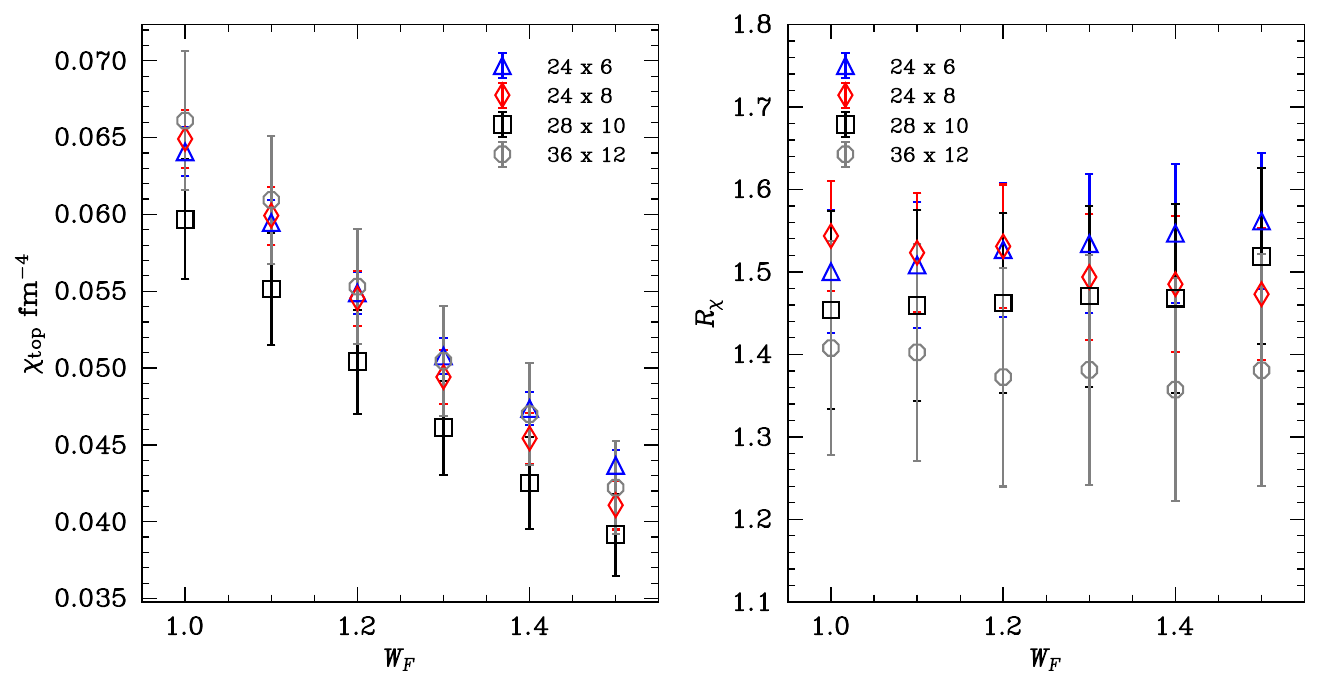}
  \caption{Reweighted topological susceptibility (\textsc{left}) and ratio of reweighted susceptibilities (\textsc{right}) as functions of the window factor for all our lattices at the lowest temperature, $T = 112$ MeV, and with a background magnetic field of $eB = 0.5$ GeV$^2$. The dependence on $W_F$ disappears when considering ratios. The improved definition of $Q_{\rm top}$ at $\tau_f/a^2 = N_t^2/8$ was used.}
  \label{window_plot}
\end{figure}

The reweighted topological susceptibility at finite lattice spacings is found to depend on $W_F$ in a non-trivial way. We show this in the left panel of Fig.~\ref{window_plot} for all of our lattices at a temperature of  $T = 112$ MeV with a magnetic field of $eB = 0.5$ GeV$^2$. This highlights the ambiguity inherent to the reweighting procedure at low temperature and the sensitivity of $\chitop$ to it. However, as we will argue, this effect is largely independent of the magnetic field, so that the reweighting becomes applicable for the ratio of topological susceptibilities at nonzero and zero $B$,
\begin{equation}
R_{\chi}(B,T)=\frac{\chitop(B,T)}{\chitop(0,T)}\,.
\end{equation}

We demonstrate that $R_\chi$ (and in particular its continuum limit) is independent of the window size in the right panel of Fig.~\ref{window_plot}. This tells us that while we cannot distinguish the topological would-be zero modes from the small eigenvalues forming the chiral condensate, the ratios of susceptibilities are insensitive to which particular eigenvalues are chosen. Hence, instead of attempting a continuum extrapolation of the susceptibility on its own, from now on we study the ratio of susceptibilities $R_{\chi}(B,T)$.

\subsection{Ratios of topological susceptibilities}

We have found that the reweighting procedure is under control for the ratios of susceptibilities and a reliable continuum extrapolation can be performed for them. These results constitute the first non-perturbative determination of the magnetic field dependence of the susceptibility on the lattice. At low temperatures, we can compare the results to chiral perturbation theory estimations, available at next-to-leading order~\cite{Adhikari:2022vqs}. Interestingly, ChPT predicts $R_{\chi}$ to behave in the same way as the similar ratio of chiral condensates~\cite{Adhikari:2022vqs}. We comment on this in Sec.~\ref{susc_lowT}, where we also investigate whether this sum rule is fulfilled beyond the region of applicability of ChPT too. At high temperatures, where the dilute instanton gas approximation is applicable and the temperature scale is much bigger than that of the magnetic field, we expect that the temperature suppression of the susceptibility overcomes any magnetic field dependence, yielding $R_{\chi} \approx 1$. However, the magnitude of the magnetic fields studied in this paper are comparable to the temperatures under consideration. Hence, a non-trivial behaviour is expected even at our highest temperature. We demonstrate that this is indeed the case in Sec.~\ref{susc_Tc}.

\subsubsection{Magnetic field dependence at low temperatures}
\label{susc_lowT}

\begin{figure}[ht]
  \centering
    \includegraphics[width=\textwidth]{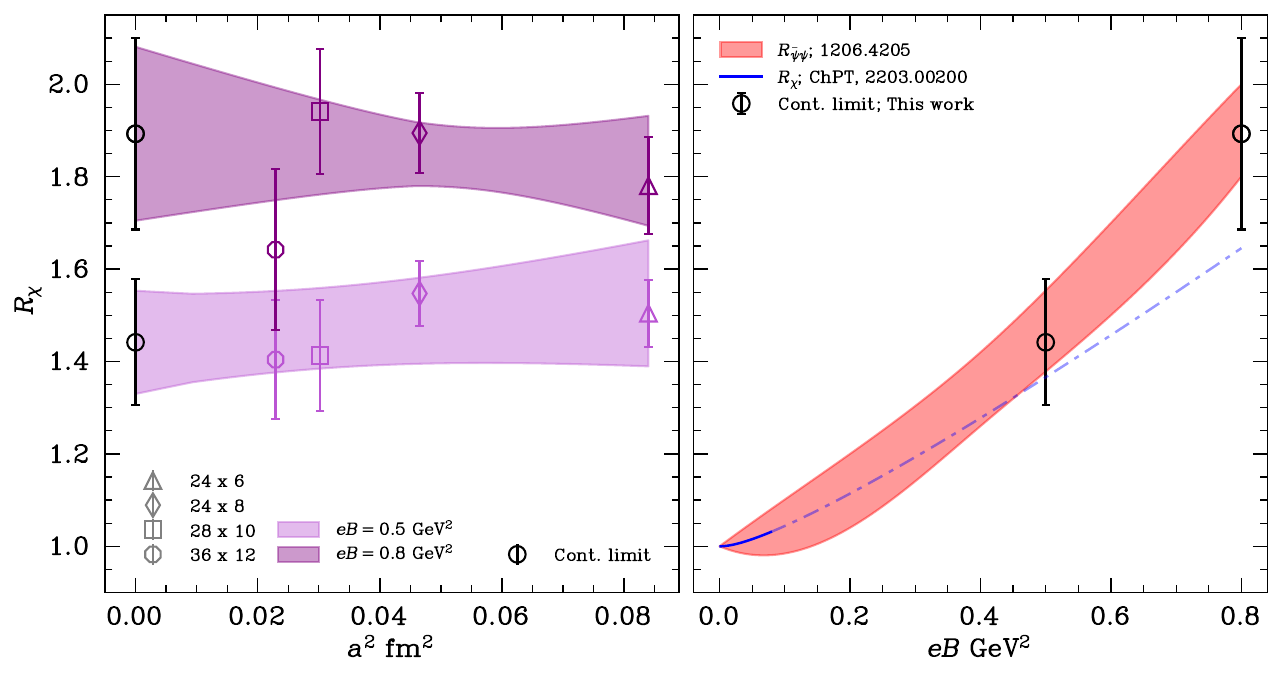}
  \caption{\textsc{Left}: Continuum extrapolations of the ratios of susceptibilities for our two magnetic fields at a temperature of 112 MeV. \textsc{Right}: Ratios of susceptibilities as a function of the magnetic field. We compared with a similar ratio of chiral condensates from~\cite{Bali:2012zg} and to a ChPT calculation~\cite{Adhikari:2022vqs}. The solid section of the blue line marks the expected region of validity of ChPT.}
  \label{clim_lowT}
\end{figure}

We begin the discussion of our main results by describing the effect of magnetic fields on the susceptibility at low temperatures. In the left panel of Fig.~\ref{clim_lowT} we show $R_\chi$ as a function of the lattice spacing for our two magnetic fields at a temperature $T = 112$ MeV. The figure includes the continuum extrapolation, which was performed assuming $\mathcal{O}(a^2)$ lattice artifacts. The systematic error was estimated by fitting with a constant function and simultaneously removing one or two data points corresponding to the coarsest lattices. Moreover, we also considered the uncertainty coming from our eight different definitions of $\Qtop$ described in Sec.~\ref{Qtop_def} by repeating the quadratic fit and adding the maximum difference in quadrature. In turn, the statistical error was computed using a jackknife procedure with data-blocking.

In the right panel we plot the result of our continuum extrapolation as a function of the magnetic field and compare it to the ChPT calculation~\cite{Adhikari:2022vqs}. Here, we also include the lattice results for the ratio $R_{\bar\psi\psi}=\langle\bar\psi\psi_{u,d}(B)\rangle/\langle\bar\psi\psi_{u,d}(0)\rangle$ of light quark condensates at finite to zero magnetic field~\cite{Bali:2012zg}, which is expected to agree with $R_\chi$ within next-to-leading-order chiral perturbation theory. We find that the ratio of susceptibilities are highly compatible not only with the ChPT prediction (especially for our smallest magnetic field) but also with the ratio of chiral condensates. Hence we show that the sum rule relating the ratios of susceptibilities and chiral condensates holds for a broad range of magnetic fields, beyond the expected region of applicability of ChPT. Our study demonstrates that the topological susceptibility can be greatly enhanced under the influence of magnetic fields, at least for $eB \leq 0.8$ GeV$^2$.

\subsubsection{Magnetic field dependence around the crossover}
\label{susc_Tc}

The continuum limit, as well as the error analysis for the remaining three temperatures was performed in a similar fashion as for the low temperature case. Subsequently, we have interpolated the results using a cubic spline fit. 

\begin{figure}[ht]
  \centering
    \includegraphics[width=0.5\textwidth]{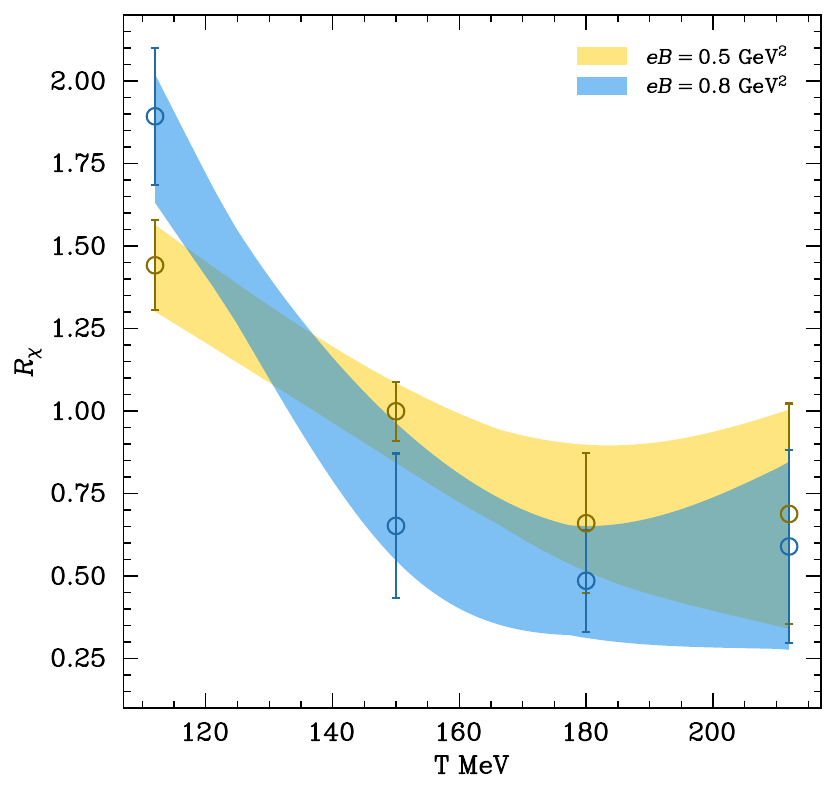}
  \caption{Continuum extrapolations of the ratios of susceptibilities for our two magnetic fields as a function of temperature.}
  \label{clim_Tdep}
\end{figure}

The final result for the temperature dependence of $R_{\chi}$ at our two magnetic fields is shown in Fig.~\ref{clim_Tdep}. First, we observe -- as discussed in the previous section -- the enhancement of the susceptibility by the magnetic field at low temperatures. This effect is smoothly suppressed as the temperature grows, leading to $R_\chi<1$ at high $T$. The rate of this suppression is found to be larger for the stronger magnetic field. In particular, at $T\approx 135$ MeV we observe that the two curves cross each other. Finally, in the high temperature range ($T = 160-212$ MeV) we observe how the suppression by the magnetic field slightly reduces. This can be interpreted as the onset of the region where the ratios begin to converge towards unity with increasing temperature. 

This behaviour of the topological susceptibility is reminiscent of that of the chiral condensate, which undergoes magnetic catalysis at low $T$ and inverse magnetic catalysis in the transition region~\cite{Bali:2012zg}. 
These results for $R_\chi$ therefore provide a new aspect of the inverse magnetic catalysis phenomenon, corroborating the relationship between the topological susceptibility and the chiral condensate, found in the previous section.

\begin{figure}[ht]
  \centering
    \includegraphics[width=\textwidth]{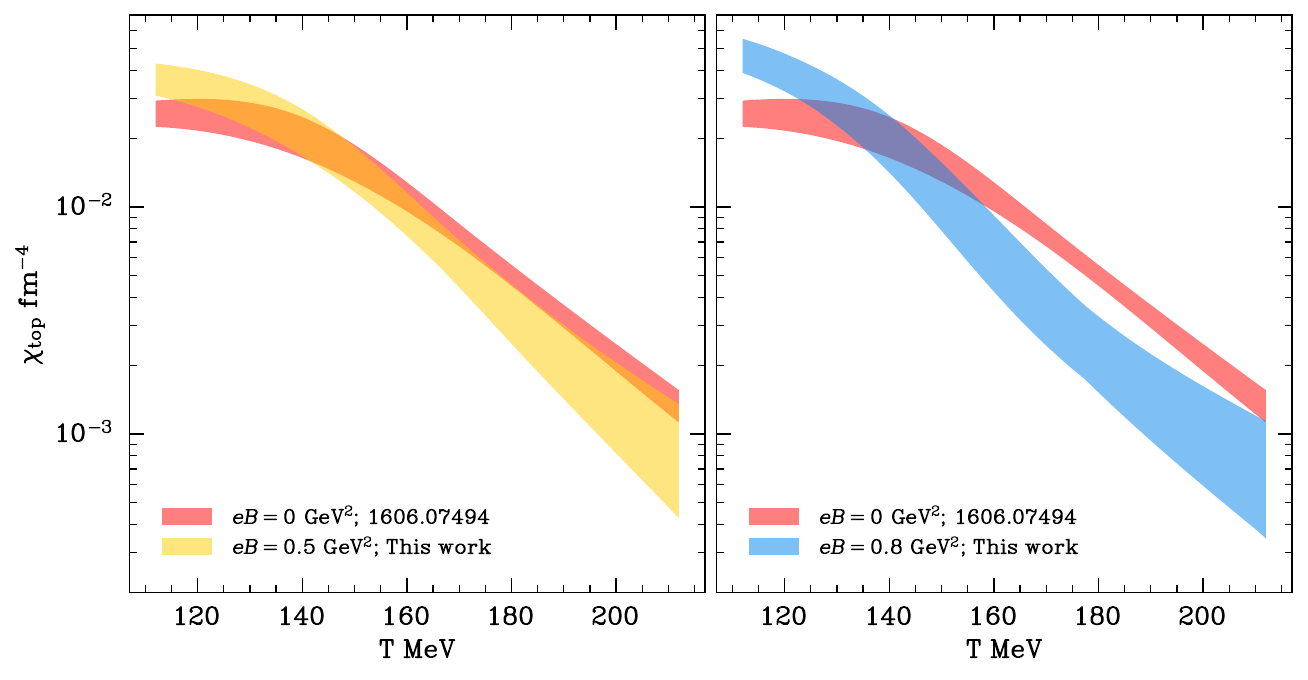}
  \caption{Comparison of the full susceptibilities as a function of temperature at a finite magnetic field (this work) to the zero magnetic field case (from~\cite{Borsanyi:2016ksw}, in red). The left and right panels show the comparisons for $eB = 0.5$ GeV$^2$ and $eB = 0.8$ GeV$^2$, respectively.}
  \label{clim_full}
\end{figure}

We also provide the behaviour of the susceptibility itself as a function of temperature, $\chi_{\rm top}(B,T)$. This was obtained by combining our results for the ratios $R_{\chi}$ with the calculation of the topological susceptibility at $B=0$ performed in~\cite{Borsanyi:2016ksw}. The topological susceptibilities as a function of both temperature and magnetic field are shown in Fig.~\ref{clim_full}. All the features previously described for the ratios (enhancement at low temperatures, inverse magnetic catalysis, suppression at higher temperatures) can be observed in this plot from a different perspective. Moreover, comparisons between our work and that of~\cite{Ali:2020jsy} in the NJL model show a qualitative agreement in the studied temperature region. In particular, both calculations show an enhancement of the susceptibility with the magnetic field at low temperatures and inverse magnetic catalysis around the crossover region.

\begin{figure}[hb]
  \centering
    \includegraphics[width=\textwidth]{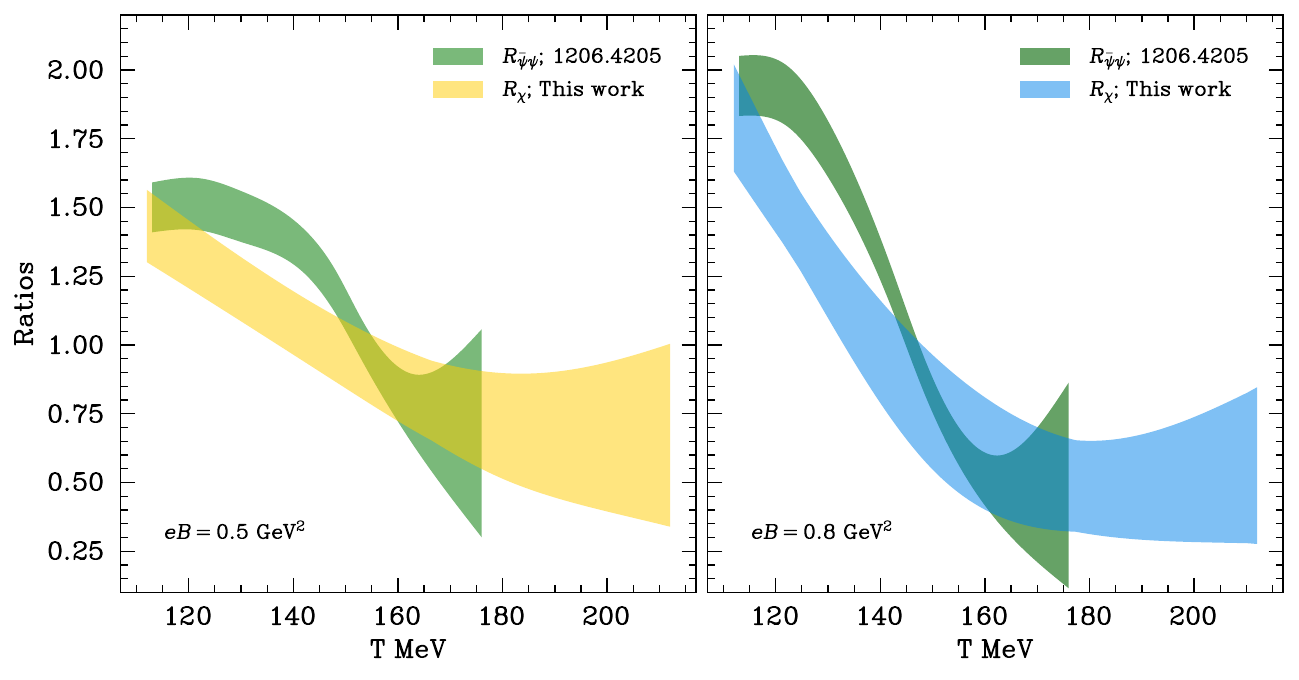}
  \caption{Comparisons of our continuum extrapolated ratios of susceptibilities $R_{\chi}$ to similar ratios of chiral condensates from~\cite{Bali:2012zg}. The left and right panels show the comparisons with $eB = 0.5$ GeV$^2$ and $eB = 0.8$ GeV$^2$, respectively.}
  \label{comp_psipsibar}
\end{figure}

Finally, and inspired by the sum rule at low temperatures, we decided to compare the ratios $R_{\bar\psi\psi}$ of chiral condensates with $R_{\chi}$ at higher temperatures. The result can be seen in Fig.~\ref{comp_psipsibar}. The sum rule is again observed to hold for low temperatures and slightly violated as the temperature is increased, according to our interpolation. As we keep increasing the temperature beyond the crossover, both ratios are found to be compatible again. The recovery of the sum rule at temperatures higher than $T_c$ is a surprising finding that has not yet been discussed in the literature so far. 

\section{Summary and outlook}
\label{sec4}

In this paper, we have performed the first non-perturbative lattice determination of the topological susceptibility in the presence of background magnetic fields (up to $eB = 0.8$ GeV$^2$) and for a broad range of temperatures around the crossover region ($T=112-212$ MeV). The simulations have been performed using 2+1 flavours of stout-smeared staggered quarks at the physical point. The gluon links have been smeared using the gradient flow, with the two-fold goal of identifying the topology and renormalising the fields.  A reweighting technique for the fermion determinant has been employed in order to reduce the lattice artifacts arising from the absence of exact zero modes in the staggered discretisation. We have shown how the ambiguity in defining the topological modes at low temperatures is avoided by considering ratios of susceptibilities, which allows us to obtain a controlled continuum extrapolation for all our temperatures and magnetic fields.

At low temperatures, we have found that the susceptibility is enhanced by the magnetic field, as predicted by ChPT, in accordance with the magnetic catalysis of the chiral condensate. Moreover, we discovered that the sum rule that relates the magnetic field dependence of ratios of chiral condensates and ratios of susceptibilities is sustained for magnetic fields much larger than those within the region of applicability of ChPT. At higher temperatures, we observe the opposite effect: for $T \gtrsim 135$ MeV, increasing the magnetic field suppresses the susceptibility -- reminiscent of the inverse magnetic catalysis of the condensate. Finally, we also compared the ratios of susceptibilities to those of chiral condensates at higher temperatures and found the two to agree quantitatively both at low $T$ and well above the crossover temperature. An interesting question is whether this sum rule is maintained for even higher temperatures. This aspect could be discussed and understood better using perturbation theory in an instanton background and magnetic fields, just like at $B=0$~\cite{Gross:1980br,Schafer:1996wv}.

In summary, our findings demonstrate the nontrivial impact of the magnetic field on the topological susceptibility and, thus, on the axion mass. Beyond the comparisons to chiral perturbation theory, the results may be used to benchmark low-energy models and effective theories of QCD. 
For comparison purposes, we attach the data necessary to reproduce all the plots on this paper as an ancillary file in {\fontfamily{qcr}\selectfont arXiv}.

\acknowledgments

The authors acknowledge support by the Deutsche Forschungsgemeinschaft (DFG, German Research Foundation) through the CRC-TR 211 `Strong-interaction matter under extreme conditions' – project number 315477589 – TRR 211 and by the Helmholtz Graduate School for Hadron and Ion Research (HGS-HIRe for FAIR). The authors are also grateful to Guy D. Moore for fruitful discussions. The computations in this work were performed on the GPU cluster at Bielefeld University. Some of the calculations were carried out using the \texttt{SIMULATeQCD} framework~\cite{Mazur:2021zgi,HotQCD:2023ghu}. 

\appendix 

\section{Parameters of the simulations}
\label{tab_parameters}

Here we tabulate the lattice parameters used in our study, including the values of the lattice spacing $a$, the gauge coupling $\beta$ and the magnetic flux $N_b$ (the first and second numbers correspond to a physical magnetic flux of $eB = 0.5$ and $0.8$ GeV$^2$, respectively). The quark masses in lattice units $am_{ud}$, $am_s$ can be obtained from the line of constant physics computed in~\cite{Borsanyi:2010cj}.

\begin{table*}[ht] \centering
\begin{small}
\begin{tabular}{cc}
\toprule
$24^3\times6$ & $24^3\times8$ \\
\begin{tabular}{ccc}
\toprule
$a$ [fm] & $\beta$ & $N_b$ \\ \midrule
0.289788 & 3.4500 & 33, 53 \\ \hdashline
0.215664 & 3.5500 & 18, 29 \\ \hdashline
0.186526 & 3.6000 & 14, 22 \\ \hdashline
0.152645 & 3.6720 & 9, 15 \\
\bottomrule
\end{tabular}&

\begin{tabular}{ccc}
\toprule
$a$ [fm] & $\beta$ & $N_b$ \\ \midrule
0.215664 & 3.5500 & 18, 29 \\ \hdashline
0.162076 & 3.6500 & 10, 17 \\ \hdashline
0.132803 & 3.7250 &  7, 11 \\ \hdashline
0.117278 & 3.7750 &  5, 9 \\ 
\bottomrule
\end{tabular}\\
\bottomrule
\end{tabular}
\end{small}
\end{table*}

\begin{table*}[ht] \centering
\begin{small}
\begin{tabular}{cc}
\toprule
$28^3\times10$ & $36^3\times12$ \\
\begin{tabular}{ccc}
\toprule
$a$ [fm] & $\beta$ & $N_b$ \\ \midrule
0.173756 & 3.6250 & 16, 26 \\ \hdashline
0.132803 & 3.7250 & 9, 15 \\ \hdashline
0.110492 & 3.8000 & 7, 10 \\ \hdashline
0.0933079 & 3.8750 & 5, 7 \\ 
\bottomrule
\end{tabular}&

\begin{tabular}{ccc}
\toprule
$a$ [fm] & $\beta$ & $N_b$ \\ \midrule
0.151415 & 3.6750 & 20, 32 \\ \hdashline
0.110492 & 3.8000 & 11, 17 \\ \hdashline
0.0913236 & 3.8850 & 7, 12 \\ \hdashline
0.0782779 & 3.9600 & 5, 9 \\ 
\bottomrule
\end{tabular}\\
\bottomrule
\end{tabular}
\caption{Parameters of the simulations in our study.}
\end{small}
\end{table*}

\section{Reweighting of the quark condensate}
\label{rw_psibarpsi}

As a further check of the validity of the reweighting technique, here we show its effect on a non-topological observable, the quark condensate $\langle\Bar{\psi}\psi_f\rangle$. On a given gluon configuration, it is defined as
\begin{equation}
    \Bar{\psi}\psi_f = \frac{1}{4}\frac{1}{\Omega}\,\mathrm{Tr}\hspace{1pt} (\slashed{D}_f+m_f)^{-1}\,,
    \label{eq:condoperator}
\end{equation}
where the $1/4$ prefactor comes from the rooting procedure. The trace of the inverse Dirac operator is calculated using noisy estimators.

Unlike the topological charge, Eq.~\eqref{eq:condoperator} is a fermionic operator that is affected directly by the reweighting procedure. In other words, the impact of reweighting is not only through the correlation of the observable with the reweighting factor $W$ from Eq.~\eqref{eq:Wdef}, but also through the modification of the operator itself. In particular, the reweighted operator reads
\begin{equation}
    \Bar{\psi}\psi_f^{\rm rw}  =  \Bar{\psi}\psi_f - \frac{1}{4}\frac{1}{\Omega}\sum_{j=1}^{2|\Qtop|}\frac{2m_f}{\lambda_{fj}^2+m_f^2} + \frac{1}{\Omega}\frac{|\Qtop|}{m_f}\,.
\end{equation}

Here, we considered the total condensate operator computed via noisy estimators (the first term), separated the contribution of would-be zero modes from it (the second term) and replaced these by the exact zero modes (the third term). For the latter two terms, we again used the eigenbasis of the Dirac operator, just like for the reweighting factor itself.

This operator will be denoted by $\bar\psi\psi_f^{\rm rw}$. The complete reweighting of the quark condensate is obtained by correlating this operator with the reweighting factor, given by $\langle\bar\psi\psi_f^{\rm rw}\rangle_{\rm rw}$. For completeness, below we also consider the impact of separately reweighting the operator or the ensemble, i.e.\ $\langle\bar\psi\psi_f^{\rm rw}\rangle$ and $\langle\bar\psi\psi_f\rangle_{\rm rw}$ as well as the unreweighted observable $\langle\bar\psi\psi_f\rangle$.

\begin{figure}[ht]
  \centering
    \includegraphics[width=\textwidth]{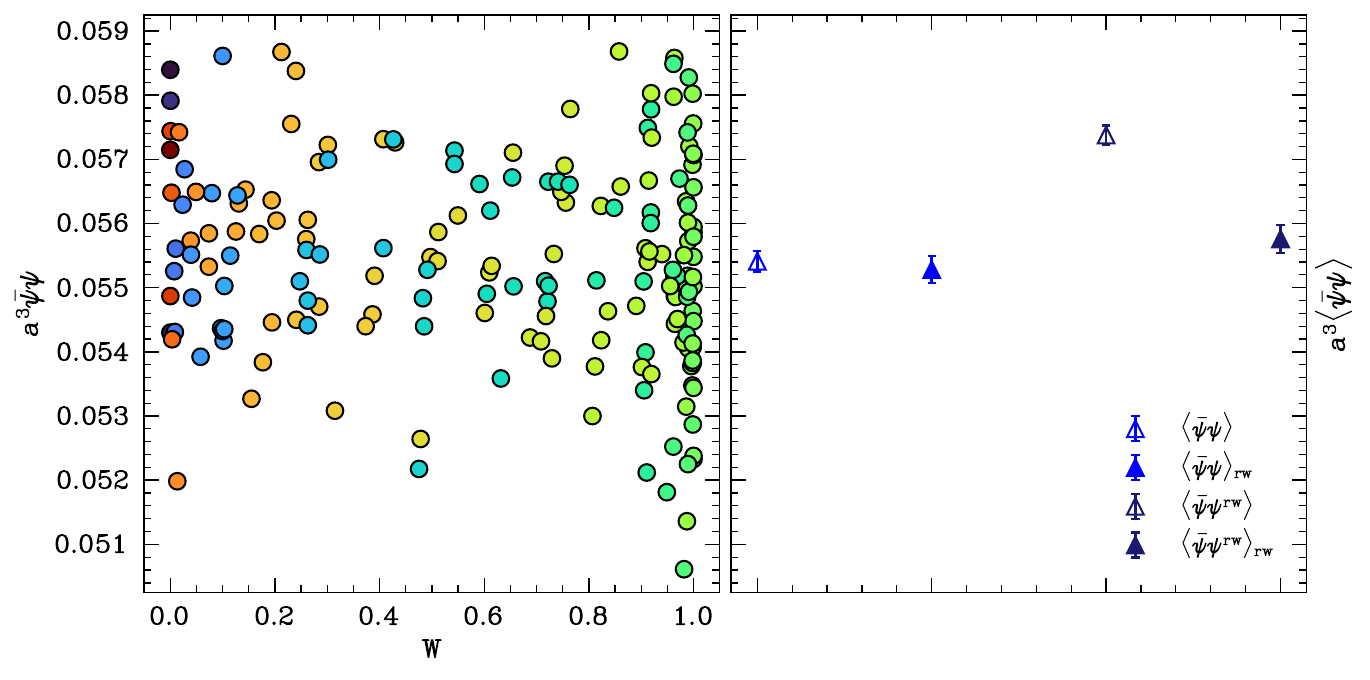}
  \caption{\textsc{Left}: Scatter plot of the reweighting factor and the quark condensate. The colour map is given by the value of $\Qtop$ (with blue denoting negative, red positive and green close to zero values). \textsc{Right}: Effect of the reweighting on the quark condensate for the operator and for the ensemble. 
  The subscript $f$ has been dropped in order to unclutter the notation. The results are for the average light quark condensate.}
  \label{psibarpsi_rw}
\end{figure}

The correlations between the reweighting factor and the quark condensate (i.e.\ between the topological charge and the condensate) are much milder than in the case of the topological susceptibility. Therefore, we expect that the prime effect of the ensemble reweighting technique will be a reduction of the effective statistics and, accordingly, higher statistical errors. This is indeed what happens, as we demonstrate in Fig.~\ref{psibarpsi_rw} using our $24^3\times 6$ ensemble at $T = 112$ MeV with zero background magnetic field for the average light quark condensate. 

The left panel of Fig.~\ref{psibarpsi_rw} shows that there is indeed no strong correlation between the reweighting factor and the quark condensate.
The corresponding expectation values are shown in the right panel. 
The ensemble reweighting merely increases the statistical error slightly. In turn, reweighting the operator enhances the quark condensate by a few percent. The completely reweighted observable is found to be compatible with the unreweighted one within one standard deviation.
These small effects are expected to disappear towards the continuum limit, as the staggered spectrum approaches the continuum one.
In summary, we conclude that the reweighting technique introduced in the main text for the topological susceptibility has no significant impact for the quark condensate.

\bibliography{biblio.bib}

\end{document}